# Computing Nature – A Network of Networks of Concurrent Information Processes

Gordana Dodig Crnkovic [1] and Raffaela Giovagnoli [2]

[1] Department of Computer Science and Networks, Mälardalen University, Sweden
    email: gordana.dodig-crnkovic@mdh.se
[2] Faculty of Philosophy. Lateran University, Rome (Italy)
    email: raffa.giovagnoli@tiscali.it

## 1   Introduction

The articles in the volume COMPUTING NATURE, forthcoming in Springer SAPERE book series http://www.springer.com/series/10087 present a selection of works from the Symposium on Natural/Unconventional Computing at AISB/IACAP (British Society for the Study of Artificial Intelligence and the Simulation of Behaviour and The International Association for Computing and Philosophy) World Congress 2012, held at the University of Birmingham, celebrating Turing centenary.

This book is about nature considered as the totality of physical existence, the universe. By physical we mean all phenomena - objects and processes - that are possible to detect either directly by our senses or via instruments. Historically, there have been many ways of describing the universe (cosmic egg, cosmic tree, theistic universe, mechanistic universe) while a particularly prominent contemporary approach is computational universe.

One of the most important pioneers of computing, Turing, described by Hodges as natural philosopher, can be identified as a forerunner and founder of the notion of computing nature and natural computing through his morphological computing and "unorganized" (neural-network type) machines. Dodig-Crnkovic and Basti in this volume address Turing's role as pioneer of natural computation.

Present day computers are distinctly different from the early stand-alone calculating machines designed to mechanize computation of mathematical functions. Computers today are networked and largely used for world-wide communication and variety of information processing and knowledge management. They are cognitive tools of extended mind used in social interactions and ever increasing repositories of information. Moreover, computers play an important role in the control of physical processes and thus connect directly to the physical world, especially in automation, traffic control and robotics. Apart from classical engineering and hard-scientific domains, computing has in recent decades pervaded new fields such as biology and social sciences, humanities and arts – all previously considered as typical soft, non-mechanical and unautomatable domains.

Computational processes running in networks of networks (such as the internet) can be modeled as distributed, reactive, agent-based and concurrent computation. The main criterion of success of this computation is not its termination, but its behavior - response to changes, its speed, generality and flexibility, adaptability, and tolerance to noise, error, faults, and damage. Internet, as well as operating systems and database management systems are designed to operate indefinitely and termination for them indicates an error. We will return to the topic of concurrent computing and its relationship with Turing machine model in more detail later on.

One of the aims of this book is to show the state of the art of developments in the field of natural/unconventional computation which can be seen as generalization and enrichment of the repertoire of classical computation models (all of them considered to be equivalent to the Turing machine model). As a generalization of the traditional algorithmic Turing Machine model of computation, (in which the computer was an isolated box provided with a suitable algorithm and an input, left alone to compute until the algorithm terminated), natural computation models interaction i.e. communication of computing processes with each other and with the environment during the computation.

Hewitt [1] characterizes the Turing machine model as an *internal* (*individual*) framework and the Actor model of concurrent computation as an *external* (*sociological*) model of computing. This tension between (isolated) individual *one* and (interacting) social *many* resonates with two articles from this volume: Cottam et al. who distinguish "conceptual umbrella of *entity* and *its ecosystem*" and Schroeder's view that "Information can be defined in terms of the categorical opposition of one and many, leading to two manifestations of information, selective and structural. These manifestations of information are dual in the sense that one always is associated with the other." Here information is directly related with computation defined as information processing. [2]

The frequent objection against the computational view of the universe, elaborated by Zenil in this volume, is that "it is hard to see how any physical system would not be computational." The next frequently asked question is: if the universe computes, what is its input and output of its computation? This presupposes that a computing system must have input from the outside and that it must deliver some output to the outside world. But actor system [1] for example needs no input. Within pancomputationalist framework, the whole universe computes its own next state from its current state [3]. As all of physics is based on quantum mechanical layers of information processing, one may say that zero-point (vacuum) oscillations can be seen as one of the constant inputs for the computational network of the universe. What causes different processes in the universe is the interaction between its parts or exchange of information. The universe is a result of evolution from the moment of big-bang or some other primordial state, through the complexification of the relationships between its actors by computation as a process of changes of its informational structure. Physical forces are established through particle exchanges (message exchanges) which necessarily connect particles into a web of physical interactions which are manifestation of natural laws. The whole of the universe is in the state of permanent flow, far from steady state, which results in forming increasingly complex structures, [4]. So much about the input-output question.

As to the objection that not all of the universe can be computational, it is essential to keep in mind the complex layered architecture of the computing nature, as not all of computation is the same – computation is proceeding on many scales, on many levels of hierarchical organization. Moreover, in tandem with computation, universe is described by information, representing its structures. *Given that computation follows physical laws*, or *represents/implements physical laws*, generative model of the universe can be devised such that some initial network of informational processes develops in time into increasingly complex (fractal, according to Kurakin) information structures.

The parallel could be drawn between natural computationalism and atomic theory of matter which implies that all of matter is made of atoms (and void). We may also say that all of physics (structures and processes) can be derived from elementary particles (and void that is an ocean of virtual particles which for short time, obeying Heisenberg uncertainty relations, pop into existence and quickly thereafter disappear). This does not make the world a soup of elementary particles where no differences can be made, and nothing new can emerge. Those basic elements can be imagined as neodymium ball magnets from which countless structures can be constructed (in space and time through interactions).

Unified theories are common and valued in physics and other sciences, and natural computationalism is such a unified framework. It is therefore not unexpected that physicists are found among the leading advocates of the new unified theory of informational and computational universe – from Wheeler, via Feynman, to our contemporaries such as Fredkin, Lloyd, Wolfram, Goyal and Chiribella. For the articles of latter two physicists on the topic of informational universe, see the special issue of the journal *Information* titled *Information and Energy/Matter* [5] and the special issue of the journal *Entropy* titled *Selected Papers from Symposium on Natural/ Unconventional Computing and its Philosophical Significance* [6].

Conceptualizing the physical world as a network of information networks evolving through processes of natural computation helps us to make more compact and coherent models of nature, connecting non-living and living worlds. It presents a suitable basis for incorporating current developments in understanding of biological, cognitive and social systems as generated by complexification of physicochemical processes through self-organization of molecules into dynamic adaptive complex systems by morphogenesis, adaptation and learning—all of which are understood as computation (information processing).

## 2    Re-Conceptualizing of Nature as Hierarchically Organized Computational Network of Networks Architecture

### 2.1    Natural Hierarchy

"If computation is understood as a physical process, if nature computes with physical bodies as objects (informational structures) and physical laws govern process of computation, then the computation necessarily appears on many different levels of organization. Natural sciences provide such a layered view of nature. One sort of computation process is found on the quantum-mechanical level of elementary particles, atoms and molecules; yet another on the level of classical physical objects. In the sphere of biology, different processes (computations = information processing) are going on in biological cells, tissues, organs, organisms, and eco-systems. Social interactions are governed by still another kind of communicative/interactive process. If we compare this to physics where specific "force carriers" are exchanged between elementary particles, here the carriers can be complex chunks of information such as molecules or sentences and the nodes (agents) might be organisms or groups—that shows the width of a difference." Dodig Crnkovic [2]

Searching for a framework for natural computation and looking at nature from variety of perspectives and levels of organization, Cottam et al. in this book address the general question of hierarchy in nature and point to Salthe who " restricts the term hierarchy to two forms: the scale (or compositional) hierarchy and the specification (or subsumption) hierarchy. However, we find that a third form – the representation or model hierarchy – is most suitable for describing the properties of Natural systems." Special emphasis is on birational ecosystemic principles: "Nature seen through sciences brings all of Science under a generalized umbrella of *entity and its ecosystem*, and then characterizes different types of entity by their relationships with their relevant ecosystems." (emphasis added)

In spite of suggested tree-structure with representation on top, followed by model hierarchy with subsequent compositional and subsumption hierarchy, the authors point out that "even here the movement between two extremes can be observed – from the bottom up and from the top down. Parts defining the whole, which once established, affect its parts. As a case in point, they provide an example of a (Natural) model hierarchy for a tree represented at different scales: "{a tree described in terms of atoms}, {a tree described in terms of molecules}, {a tree described in terms of cells}… up to {a tree described in terms of branches}, {a tree as itself – a tree}". Here inter-scale interfacing and consequently digital-analog interfaces are discussed and it is emphasized that naturally-hierarchical multi-scale organisms function differently from a digital computer. The article concludes with the hope that this birational ecosystemic hierarchical framework will be capable of defining computation which is closer to processes that we find in nature.

### 2.2 Cognitive Level of Information Processing

In a hierarchy of organizational levels in nature the most complex level of information processing is cognitive level and it subsumes all lower levels that successively emerge from their antecedent lower levels. Lindley in this volume addresses the problems encountered in the development of engineered autonomous and intelligent systems caused by *exclusively linguistic models of intelligence*. The alternative proposed is "taking inspiration more directly from biological nervous systems". This approach is argued to be able to go "far beyond twentieth century models of artificial neural networks (ANNs), which greatly oversimplified brain and neural functions". This implies study of computation as information processing in neural and glial systems in order to *implement* "asynchronous, analog and self-* architectures that digital computers can only *simulate*." (emphasis added)

Continuing on the level of neural systems, Phillips's paper addresses the important topic of coordination of concurrent probabilistic inference. Adaptively organized complexity of life builds on information processing and in cognitive agents with neural systems also on inference. The paper discusses the theory of Coherent Infomax in relation to the Theory of free energy reduction of probabilistic inference. Coherent Infomax show how neural systems combine local reliability with context-sensitivity and here we recognize the leitmotif of individual and social, agent and its eco-system from several other papers.

Bull et al. in this volume address Turing's unorganized machines as models of neural systems - a form of discrete dynamical system. Turing in his 1948 paper [7] made an essential insight about the connection of social aspects of learning and intelligence. From the contemporary perspective of natural computing we see networks as information processing mechanisms and their role in intelligence is fundamental. Suggesting that natural evolution may provide inspiration for search mechanisms to design machines, Bull et al investigate Turing's dynamical representation for networks of vesicles, membrane-bound compartments filled with Belousov-Zhabotinsky chemical mixture, used as liquid information processing system. Communication between vesicles was implemented as chemical signals - excitations propagating between vesicles, which was seen as imitation or cultural information communication, that may provide "a useful representation scheme for unconventional computing substrates".

Even Arriola-Rios et al. contribute to this book with a high level of abstraction, interdisciplinary perspective, this time on object representation in animals and robots from segregate information about objects. This work makes contribution to better understanding of the details of information processing on cognitive level. As information in a cognizing agent forms internal representations, which depend on the way of agent's use of information, it could be compressed and thus re-used "for deriving interpretations, causal relationships, functions or affordances". Particular analysis is devoted reasoning about deformable objects.

## 3   The Unreasonable Effectiveness of Mathematics in the Natural Sciences (Except for Biology). Mathematicians Bias and Computing Beyond the Turing Limit

Mathematician's contribution to the development of the idea of computing nature is central. Turing as an early proponent of natural computing put forward a machine model that is still in use. How far can we hope to go with Turing machine model of computation?

In the context of computing nature, living systems are of extraordinary importance as up to now science haven't been able to model and simulate the behavior of even the simplest living organisms. "The unreasonable effectiveness of mathematics" ob-served in physics (Wigner) is missing for complex phenomena like biology that today lack mathematical effectiveness (Gelfand), see [8].

Not many people today would claim that human cognition (information processing going on in our body, including thinking) can be adequately modeled as a result of computation of one Turing machine, however complex function it might compute. In the next attempt, one may imagine a complex architecture of Turing machines running in parallel as communicating sequential processes (CSPs) exchanging information. We know today that such a system of Turing machines cannot produce the most general kind of computation, as truly asynchronous concurrent information processing going on in our brains.

However, one may object that IBM's super-computer Watson, the winner in man vs. machine "Jeopardy!" challenge, runs on contemporary (super)computer which is claimed to be implementations of Turing machine. Yet, Watson is connected to the Internet. And Internet is not a Turing machine equivalent. It is not even a network of Turing Machines. Information processing going on throughout the entire Internet includes signaling and communication based on complex asynchronous physical processes that cannot be sequentionalized. (Hewitt, Sloman) As an illustration see Barabási et al. article [9] on parasitic computing that *implements computation on the communication infrastructure of the internet, thus using communication for computation.*

Zenil in this volume examines the question: "What does it mean to claim that a physical or natural system computes?" He proposes a behavioural characterisation of computing in terms of a measure of programmability, which reflects a system's ability to react to external stimuli. To that end Zenil investigates classical foundations for unconventional computation.

Hernandez-Espinosa and Hernandez-Quiroz, starting from the old computationalism defined as the claim that the human mind can be modeled by Turing Machines, analyze Wolfram's Principle of Computational Equivalence – the claim that "any natural (and even human) phenomenon can be explained as the interaction of very simple rules."

The next step would be to replace present basic simple rules of cellular automata with more elaborate ones. Instead of synchronous update of the the whole system; they can be made asynchronous networks of agents, placed in layered architectures on different scales etc. The basic idea of generative science is to generate apparently unanticipated and infinite behaviour based on deterministic and finite rules and parameters reproducing or resembling the behavior of natural and social phenomena. As an illustration see Epstein, [10].

If we want to generalize the idea of computation to be able to encompass more complex operations than mechanical execution of an algorithm, simulating not only a person executing strictly mechanical procedure, but the one constructing new theory, we must go back to underlying mathematics.

Burgin and Dodig-Crnkovic, in present volume analyze methodological and philosophical implications of algorithmic aspects of unconventional/natural computation that extends the closed classical universe of computation of the Turing machine type. The new model constitute an open world of algorithmic constellations, allowing in-creased flexibility and expressive power, supporting constructivism and creativity in mathematical modeling and enabling richer understanding of computation, see [11].

### 3.1 Hypercomputation - Beyond the Turing Limit

Hypercomputation is the research field that formulated the first ideas about the possibility of computing beyond Turing machine model limits. The term hypercomputation was introduced by Copeland and Proudfoot [12]. The expression "super-Turing computation" was coined by Siegelman and usually implies that the model is supposed to be physically realizable, while hypercomputation frequently relies on thought experiments. Present volume offers two contributions that sort under hypercomputation, written by Franchette and Douglas.

Franchette studies the possibility of a physical device that hypercomputes by building an oracle hypermachine, "namely a device which is to be able to use some extern information from nature to go beyond Turing machines limits." The author addresses an analysis of the verification problem for oracle hypermachines.

Douglas in his contribution presents a critical analysis of Siegelmann Networks.

### 3.2 Physical Computation "In Materio" - Beyond the Turing Limit

Several authors at the Symposium on Natural/Unconventional Computing at AISB/IACAP World Congress 2012 (Stepney, Cooper, Goyal, Basti, Dodig-Crnkovic) emphasized the importance of physical computing, or as Stepney [13] termed it, "computation in materio".

Barry Cooper: What Makes A Computation Unconventional? (coming paper)

Cooper in his article Turing's Titanic Machine? [14] diagnoses the limitations of the Turing machine model and identifies the ways of overcoming those limitations:

- Embodiment invalidating the `machine as data' and universality paradigm.
- The organic linking of mechanics and emergent outcomes delivering a clearer model of supervenience of mentality on brain functionality, and a reconciliation of different levels of effectivity.
- A reaffirmation of the importance of experiment and evolving hardware, for both AI and extended computing generally.
- The validating of a route to creation of new information through interaction and emergence.

Related article by the same author elucidates the role of physical computation vs universal symbol manipulation: The Mathematician's Bias and the Return to Embodied Computation from the book A Computable Universe: Understanding and Exploring Nature As Computation. [15]

The theme of embodied computation is addressed in this volume by Hernandez-Quiroz and Padilla who examine actual physical realizability of mathematical constructions of abstract entities - a controversial issue and important in the debate about limits of Turing model. The authors study a simple special of "physical realizability of the enumeration of rational numbers by Cantor's diagonalization by means of an Ising system".

### 3.3 Higher Order Computability - Beyond the Turing Limit

One of the main steps towards the new paradigm of natural/unconventional computing is to make visible host of myths surrounding the old paradigm and helping it to survive. One of those myths is that our modern computers with all their programming languages are diverse implementations of Turing machines. However, as Kanneganti and Cartwright already argued twenty years ago:

> "Classic recursion theory asserts that all conventional programming languages are equally expressive because they can define all partial recursive functions over the natural numbers. This statement is misleading because programming languages support and enforce a more abstract view of data than bit strings. In particular, most real programming languages support some form of higher-order data such as potentially infinite streams (input and output), lazy trees, and functions." Kanneganti and Cartwright [16]

Kleene was a pioneer of higher order computability as he "opened the frontiers of computability on higher type objects in a series of papers first on constructive ordinals and hierarchies of number-theoretical predicates and later on computability in higher types. "The form of the changing surface of the stream appears to constrain the movements of the molecules of water, while at the same time being traceable back to those same movements." Soare [17]

Cooper [18] underlines the importance of higher-order computational structures as characteristic of human thinking. This can be connected to higher-order functional programming, which means, among others, programming with functions whose input and/or output may consist of other functions.

> "Kreisel [21] was one of the first to separate cooperative phenomena (not known to have Turing computable behaviour), from classical systems and proposed [22] (p˙143, Note 2) a collision problem related to the 3-body problem as a possible source of incomputability, suggesting that this might result in "*an analog computation of a non-recursive function (by repeating collision experiments sufficiently often)*". This was before the huge growth in the attention given to chaos theory, with its multitude of different examples of the generation of informational complexity via very simple rules, accompanied by the emergence of new regularities (see for example the two classic papers of Robert Shaw [33], [32]). We now have a much better understanding of the relationship between emergence and cha-os, but this still does not provide the basis for a practically computable relationship." Cooper [18] (emphasis added)

## 4     Concurrent Computing and Turing Machine Model

### 4.1    Bi-Directional Model Development of Natural Computation

Turing machine model (originally named "Logical calculating machine") was developed by Turing in order to describe a human (at that time called "computer") executing an algorithm:

> "It is possible to produce the effect of a computing machine by writing down a set of rules of procedure and asking a man to carry them out. Such a combination of a man with written instructions will be called a 'Paper Machine'. A man provided with paper, pencil, and rubber, and subject to strict discipline, is in effect a universal machine." Turing [7]

The underlying logic of Turing's "logical calculating machine" is fully consistent standard logic. Turing machine is assumed always to be in a well defined state. [1]

In contemporary computing machinery, however, we face both states that are not well defined (in the process of transition) and states that contain inconsistency:

> "Consider a computer which stores a large amount of information. While the computer stores the information, it is also used to operate on it, and, crucially, to infer from it. Now it is quite common for the computer to contain inconsistent information, because of mistakes by the data entry operators or because of multiple sourcing. This is certainly a problem for database operations with theorem-provers, and so has drawn much attention from computer scientists. Techniques for removing inconsistent information have been investigated. Yet all have limited applicability, and, in any case, are not guaranteed to produce consistency. (There is no algorithm for logical falsehood.) Hence, even if steps are taken to get rid of contradictions when they are found, an underlying paraconsistent logic is desirable if hidden contradictions are not to generate spurious answers to queries." Priest and Tanaka [19]

Open, interactive and asynchronous systems have special requirements on logic. Goldin and Wegner [20], and Hewitt [1] argue e.g. that computational logic must be able to model interactive computation, and that classical logic must be robust towards inconsistencies i.e. must be paraconsistent due to the incompleteness of interaction.

As Sloman [21] points out, concurrent and synchronized machines are equivalent to sequential machines, but some concurrent machines are asynchronous, and thus not equivalent to Turing machines. If a machine is composed of asynchronous concurrently running subsystems, and their relative frequencies vary randomly, then such a machine cannot be adequately modeled by Turing machine, see also [3].

Turing machines are discrete but can in principle approximate machines with continuous changes, yet cannot implement them exactly. Continuous systems with nonlinear feedback loops may be chaotic and impossible to approximate discretely, even over short time scales, see [22] and [1]. Clearly Turing machine model of computation is an abstraction and idealization. In general, instead of idealized, symbol-manipulating models, more and more physics-inspired modeling is taking place.

Theoretical model of concurrent (interactive) computing corresponding to Turing machine model of algorithmic computing is under development. (Abramsky, Hewitt, Wegner) From the experience with present day networked concurrent computation it becomes obvious that Turing machine model can be seen as a special case of a more general computation. During the process of learning from nature how to compute, we both develop computing and at the same time improve understanding of natural phenomena.

> "In particular, the quantum informatic endeavor is not just a matter of feeding physical theory into the general field of natural computation, but also one of using high-level methods developed in Computer Science to improve on the quantum physical formalism itself, and the understanding thereof. We highlight a seemingly contradictory phenomenon: passing to an abstract, categorical quantum informatics formalism leads directly to a simple and elegant graphical formulation of quantum theory itself, which for example makes the design of some important quantum informatic protocols completely transparent. It turns out that essentially all of the quantum informatic machinery can be recovered from this graphical calculus. But in turn, this graphical formalism provides a bridge between methods of logic and computer science, and some of the most exciting developments in the mathematics of the past two decades" Abramsky and Coecke [23]

The similar two-way process of learning is visible in biocomputing, see Rozenberg and Kari [24]. As we already mentioned "the unreasonable effectiveness of mathematics in the natural sciences" does not (yet) apply to biology, as modeling of biological systems attempted up to now was too crude. Living systems are essentially open and in constant communication with the environment. New computational models must be interactive, concurrent, and asynchronous in order to be applicable to biological and social phenomena and to approach richness of their information processing repertoire.

Present account of models of computation highlights several topics of importance for the development of new understanding of computing and its role: natural computation and the relationship between the model and physical implementation, interactivity as fundamental for computational modeling of concurrent information processing systems such as living organisms and their networks, and new developments in logic needed to support this generalized framework. Computing understood as information processing is closely related to natural sciences; it helps us recognize connections between sciences, and provides a unified approach for modeling and simulating of both living and non-living systems. [3]

## 4.2 Concurrency vs. Symbolic Computation of Function Values

In his article: What is computation? Concurrency versus Turing's Model, Hewitt [1] makes the following very apt analysis of the relationship between Turing machines and concurrent computing processes:

> "Concurrency is of crucial importance to the science and engineering of computation in part because of the rise of the Internet and many-core architectures. However, concurrency extends computation beyond the conceptual framework of Church, Gandy [1980], Gödel, Herbrand, Kleene [1987], Post, Rosser, Sieg [2008], Turing, etc. *because there are effective computations that cannot be performed by Turing Machines.* In the Actor model [Hewitt, Bishop and Steiger 1973; Hewitt 2010], computation is conceived as distributed in space where computational devices communicate asynchronously and the entire computation is not in any well-defined state. (An Actor can have information about other Actors that it has received in a message about what it was like when the message was sent.) Turing's Model is a special case of the Actor Model." Hewitt [1] (emphasis added)

According to natural computationalism/pancomputationalism [3] every physical system is computational, but there are many different sorts of computations going on in nature seen as a network of agents/actors exchanging "messages". The simplest agents communicate with simplest messages such as elementary particles (with 12 kinds of matter and 12 anti-matter particles) exchanging 12 kinds of force-communicating particles. An example from physics is Yukawa's theory of strong nuclear force understood as exchange of mesons, which explained the interaction between nucleons. Complex agents like humans communicate through languages which use very complex messages for communication. Natural computational systems as networks of agents exchanging messages are in general asynchronous concurrent systems. Conceptually, agent-based models and Actor model are closely related, and as mentioned, even the understanding of interactions (forces) between elementary particles as exchanges of elementary particles fits in this framework.

## 4.3 Physical Computing - New Computationalism.
## Embodied Networks and Symbolic Representation

It is often argued that computationalism is the opposite of connectionism and that connectionist networks and dynamic systems do not compute. However, if we define computation in a more general sense of natural computation, instead of high level symbol manipulation of Turing machine, it is obvious that connectionist networks and dynamical systems do compute. That is also the claim made by Scheutz in the Epilogue of the book Computationalism: New Directions [25], where he notices that:

> "Today it seems clear, for example, that classical notions of computation alone cannot serve as foundations for a viable theory of the mind, especially in light of the real-world, realtime, embedded, embodied, situated, and interactive nature of minds, although they may well be adequate for a limited subset of mental processes (e.g., processes that participate in solving mathematical problems). Reservations about the classical conception of computation, however, do not automatically transfer and apply to real-world computing systems. This fact is often ignored by opponents of computationalism, who construe the underlying notion of computation as that of Turing-machine computation." Scheutz [25] p. 176

Classical computationalism was the view that classical theory of computation (Turing-machine-based, universal, disembodied) might be enough to explain cognitive phenomena. New computationalism (natural computationalism) emphasizes that embodiment is essential and thus physical computation - natural computationalism.

The view of Scheutz is supported by O'Brien [26] who refers to Horgan and Tienson [27] arguing that "cognitive processes, are not governed by exceptionless, representation-level rules; they are instead the work of defeasible cognitive tendencies subserved by the non-linear dynamics of the brains neural networks."

Dynamical characterization of the brain is consistent with the analog interpretation of connectionism. But dynamical systems theory is often not considered to be a computational framework. O'Brien [26] notices that *"In this sense, dynamical systems theory dissolves the distinction between intelligent and unintelligent behaviour, and hence is quite incapable, without supplementation, of explaining cognition. In order for dynamical engines to be capable of driving intelligent behaviour they must do some computational work: they must learn to behave as if they were semantic engines."*

O'Brien and Opie [28] thus search for an answer to the question how connectionist networks compute, and come with the following problem characterization:

"Connectionism was first considered as the opposed to the classical computational theory of mind. Yet, it is still considered by many that a satisfactory account of how connectionist networks compute is lacking. In recent years networks were much in focus and agent models as well so the number of those who cannot imagine computational networks has rapidly decreased. Doubt about computational nature of connectionism frequently takes the following two forms.

1. "(W)hile connectionists typically interpret the states and activity of connectionist networks in representational terms, closer scrutiny reveals that these putative representations fail to do any explanatory work, and since there is ''no computation without representation'' (Pylyshyn 1984, p. 62), the connectionist framework is better interpreted non-computationally.

2. Moreover it is argued that "the connectionist networks are better characterized as dynamical systems rather than computational devices."

In the above denial of connectionist models computational nature the following confusions are evident.

1. Even though it is correct that there is "no computation without representation", representation in this context can be any step in the process of information transformation from the physical world (object) to the cognitive state where an agent "recognizes" the object represented. It can be the dynamical state induced in the agents' brain as a consequence of perception of an object.

2. Dynamical systems compute as well and their computation in general is natural computation. One of the central questions in this context is the distinction between symbolic and non-symbolic computing. Trenholme [29] describes the relationship of analog vs. symbolic simulation:

> "Symbolic simulation is thus a two-stage affair: first the mapping of inference structure of the theory onto hardware states which defines symbolic computation; second, the mapping of inference structure of the theory onto hardware states which (under appropriate conditions) qualifies the processing as a symbolic simulation. *Analog simulation, in contrast, is defined by a single mapping from causal relations among elements of the simulation to causal relations among elements of the simulated phenomenon*." Trenholme [29] p.119.

Both symbolic and sub-symbolic (analog) simulations depend on causal/analog/physical and symbolic type of computation on some level but in the case of symbolic computation it is the symbolic level where information processing is observed. Similarly, even though in the analog model symbolic representation exists at some high level of abstraction, it is the physical agency and its causal structure that define computation (simulation).

Basti in this volume suggests how to "integrate in one only formalism the physical ("natural") realm, with the *logical-mathematical* ("computation") one, as well as their relationships. That is, the passage from the realm of the *causal* necessity ("natural") of the physical processes, to the realm of the *logical* necessity ("computational"), eventually representing them either in a sub-symbolic, or in a symbolic form. This foundational task can be performed, by the newborn discipline of *theoretical formal ontology*." Proposed ontology is based on the information-theoretic approach in quantum physics and cosmology, the information-theoretic approach of dissipative QFT (Quantum Field Theory) and the theoretical cognitive science.

Freeman [30] characterizes accurately the relationship between physical/sub-symbolic and logical/symbolic level in the following:

> "Human brains intentionally direct the body to make symbols, and they use the symbols to represent internal states. *The symbols are outside the brain. Inside the brains, the construction is effected by spatiotemporal patterns of neural activity that are operators, not symbols.* The operations include formation of sequences of neural activity patterns that we observe by their electrical signs. *The process is by neurodynamics, not by logical rule-driven symbol manipulation.* The aim of simu-lating human natural computing should be to simulate the operators. In its simplest form natural computing serves for communication of meaning. Neural operators implement non-symbolic communication of internal states by all mammals, including humans, through intentional actions. (…) *I propose that symbol-making operators evolved from neural mechanisms of intentional action by modification of non-symbolic operators*." (emphasis added)

Consequently, our brains use non-symbolic computing internally in order to manipulate relevant external symbols!

So in what way is physical computation/natural computation important vis-à-vis Turing machine model? One of the central questions within computing, cognitive science, AI and other related fields is about computational modeling (and simulating) of intelligent behaviour. What can be computed and how? It has become obvious that we must have richer models of computation, beyond Turing machines, if we are to efficiently model and simulate biological systems. What exactly can we learn from nature and especially from intelligent organisms?

It has taken more than sixty years from the first proposal of Turing test he called the "Imitation Game", described in Turing [31] p. 442, to the Watson machine winning Jeopardy. That is just the beginning of what Turing believed one day will be possible - a construction of computational machines capable of generally intelligent behavior as well as the accurate computational modeling of the natural world. So there are several classes of problems that deserve our attention when talking about computing nature.

To "*compute*" nature by any kind of computational means, is to model and/or simulate the behaviors of natural systems by computational means. Watson is a good example. We know that we do not function like Watson or like chess playing programs that take advantage of brute force algorithms to search the space of possible states. We use our experience, "gut feeling" and "fingertip-feeling"/ "fingerspitzengefühl" that can be understood as embodied, physical, sub-symbolic information processing mechanisms we acquire by experience and use when necessary as automatized hardware-based recognition tools.

To compute "*nature*" means to interpret natural processes, structures and objects as a result of natural computation which is in general defined as information processing. This implies understanding and modeling of physical agents, starting from the very fundamental level via several emergent levels of chemistry, biology, cognition and extended cognition (social and augmented by computational machinery).

At the moment we have bits and pieces of the picture – COMPUTING nature, that is computational modeling of nature and computing NATURE, that is nature understood in itself as a computational network of networks.

## 5 The Relationship Between Human Representation, Animal Representation and Machine Representation

We would like to highlight the relevance of the relationship between human representation and machine representation to show the main issues concerning "functionalism" and "connectionism". We propose to discuss the notion of "representation" because an important challenge for AI is to simulate not only the "phonemic" and "syntactic" aspects of mental representation but also the "semantic" aspect. Traditionally, philosophers use the notion of "intentionality" to describe the representational nature of mental states namely intentional states are those that "represent" something, because mind is directed toward objects. We think that it is important to consider the relevance of "embodied cognition" for contentful mental states (see, for instance, the classical thought experiment of the "Chinese room" introduced by Searle to criticize the important results of the Turing test, [32]).

The challenge for AI is therefore to approximate to human representations i.e. to the semantic content of human mental states. There are two competing interpretations of mental representations relevant for AI. The first focuses on the discreteness of mental representations and the second focuses on their inter-relation [33]. The first corresponds to the symbolic paradigm in AI, according to which mental representations are symbols. Proponents of the symbolic representation point on a semantic that rests on the relation between tokens of the symbol and objects of representation. The intentional mechanism functions in a way that the content of a symbol does not depend on the content of other symbols. In this sense, each symbol is discretely conferred with its intentional content. The second corresponds to connectionism in AI, according to which mental representations are distributed patterns. Proponents of this view intend the way in which a mental representation is conferred with its intentional content as mediated by relations with other representations. The virtue of connectionism as presented in the neural networks resides in the fact that the categories represented admit borderline cases of membership. As regards the composition of mental representations, it reveals itself to be the complex, contextually modulated interaction of patterns of activation in a highly interconnected network. We aim to describe the main aspects of the two approaches to make clear: the mechanisms characterizing the different way by which representations are conferred with their intentional content; the nature and structure of the categories represented and the ways in which mental representations interact.

The task to consider the similarity between human and artificial representation could involve the risk of skepticism about the possibility of "computing" this mental capacity. If we consider computationalism as defined in purely abstract syntactic terms then we are tempted to abandon it because human representation involves "real world constrains". But, a new view of computationalism could be introduced that takes into consideration the limits of the classical notion and aims at providing a concrete, embodied, interactive and intentional foundation for a more realistic theory of mind [25]. We would like to highlight also an important and recent debate on "digital representation" [34] that focus on the nature of representations in the computational theory of mind (or computationalism). The starting point is the nature of mental rep-resentations, and, particularly, if they are "material". There are authors such as Clark who maintain that mental representation are material [35] while others like Speaks think that thought processes use conventional linguistic symbols [36]. The question of digital representation involves the "problem of physical computation" [37] as well as the necessity of the notion of representation [38] so that we only have the problem of how to intend the very notion of representation [39, 40]. But, there is also the possibility of understanding computation as a purely physical procedure where physical objects are symbols processed by physical laws on different levels of organization that include "every natural process" in a "computing universe" [41]. In this context, we need a plausible relation between computation and information. Info-computational naturalism describes the informational structure of the nature i.e. a succession of level of organization of information. Morphology is the central idea in the understanding of the connection between computation and information. It proceeds by abstracting the principles via information self-structuring and sensory-motor coordination. The sensory-motor coordination provides an "embodied" interaction with the environment: information structure is induced in the sensory data, thus facilitating perception, learning and categorization.

Among the possibilities to compute human representational processes, Basti in this volume proposes a natural account from the field of formal ontology. In particular, he implements the so-called "causal theory of reference" in dynamic systems.

We think it is necessary to find a plausible philosophical strategy to consider the capacities that are common to human and machine representation (Giovagnoli in this volume). Analytic Pragmatism that is represented by the American philosopher Brandom [42] suggests relevant ideas to describe human, animal and artificial capacities for representing the external world. It is easier to start with the human case and so to describe discursive practices and to introduce norms for deploying an autonomous vocabulary, namely a vocabulary of a social practice (science, religion etc.). These norms are logical and are at the basis of an "inferential" notion of representation. But, inference in this sense, recalling Frege, is material. Brandom refuses the explanation of representation in terms of syntactical operations as presented by "functionalism" in "strong" artificial intelligence (AI or GOFAI). He does not even accept weak AI (Searle), rather he aims to present a "logical functionalism" characterizing his analytic pragmatism. According to Brandom, we are not only creatures who possess abilities such as to respond to environmental stimuli we share with thermostats and parrots but also "conceptual creatures" i.e. we are logical creatures in a peculiar way and we need a plausible view to approach human capacities.

Very interesting results are offered by Arriola-Rios and Demery et al. who discuss in this book how salient features of objects can be used to generate compact representations in animals and robots, later allowing for relatively accurate reconstructions and reasoning. They would like to propose that when exploration of objects occurs for forming representations, it is not always random, but also structured, selected and sensitive to particular features and salient categorical stimuli of the environment. They introduce how studies into artificial agents and into natural agents are complementary by emphasizing some findings from each field.

Along this line, Bull, Holley, De Lacy Costello and Adamatzky present initial results from consideration of using Turing's dynamical representation within unconventional substrate – networks of Belousov-Zhabotinsky vehicles – designed by an imitation based i.e. cultural approach. Over sixty years ago, Alan Turing presented a simple representation scheme for machine intelligence namely a discrete dynamical system network of two-input NAND gates. Since then only a few other explorations of these unorganized machines are known. As the authors underscore in their paper, it has long been argued that dynamic representations provide numerous useful features, such as an inherent robustness to faults and memory capabilities by exploiting the structure of their basins of attraction:

> "For example, unique attractors can be assigned to individual system states/outputs and the map of internal states to those attractors can be constructed such that multiple paths of similar states lead to the same attractor. In this way, some variance in the actual path taken through states can be varied, e.g., due to errors, with the system still responding appropriately. Turing appears to have been thinking along these lines also".

## 6    Conclusions, Open Problems and Future Work

*"It turns out to be better to use the world as its own model." Brooks* [43]

As already argued, we enjoy and appreciate what Wigner named "*the unreasonable efficiency of mathematics in natural sciences*" [44] – except for biology. Time is right to address biology at last and try to find out how best to use computation to model and simulate behavior of biological systems. In this context it can never be overemphasized that:*"nothing in biology makes sense except in the light of evolution"* – an insight made by the evolutionary biologist Dobzhansky [45]. In order to model (simulate) evolution we need generative models (as demonstrated by e.g. Epstein [10] and Wolfram [46]), capable of producing complex behaviors starting from simple structures and processes (rules).

Of all biological phenomena, cognition (the ability of living organisms to process information beyond simple reactivity) seems to be the most puzzling one, as in more complex organisms it is related to complex phenomena such as mind, intelligence and mental (thought) processes. Cognition in highly developed organisms indeed looks like a miracle if one does not take into account that it took enormous time in nature to develop, in the process of evolution, through the variety of biological structures and processes.

For the future work it remains to reconstruct the details of evolution of life in terms of information and computation. This especially goes for the evolution of nervous system and brains in animals and thus the development of complex cognitive capacities (such as intelligence). This understanding of the evolution and development of organisms in terms of information and computation will lead to improved understanding of underlying mechanisms of morphological computing as information self-structuring [47] Through the reverse engineering of evolution of various capacities in organisms we will be able both to deeper understand how organisms function through the detailed computational models and simulations (such as in *Human Brain/Blue Brain project*). At the same time we will learn to compute in novel and more powerful ways (such as in IBM's project of *Cognitive Computing*).

Here is the list of some important questions to answer in the framework of natural computation (information processing in physical systems).

*Generative modeling of the evolution and development of physical structures of the universe,* starting with minimum assumptions about primordial universe in terms of information and computation, based on actor (agent) networks exchanging information (messages).

*Generative modeling of hierarchical structure of emergent layers of organization* in physical systems in terms of natural computing. Modeling of the process in which the whole constraints its parts and showing how its (higher level) properties emerge.

Using natural computation to program nano-devices and applying it to universally programmable intelligent matter. (MacLennan)[48]

*Understanding and describing of the evolution and development of living organisms* on earth within the framework of natural computation (morphological computation, self-organization of informational structures through computational processes – concurrent computational processes, modeled as above.

*Understanding intelligence and consciousness,* in terms of information and computation. Explaining how representations (symbolic level) emerge from sub symbolic information processes. Understanding how exactly our brains process information, learn and act in terms of information and natural computation on different levels of organization. Working out the connections between connectionist networks/dynamic systems and symbol manipulation, sub-symbolic and symbolic information processing.

*Explaining how physics connects to life* and how the fact that we evolved from physical matter defines the ways we interact with the universe and form our concepts and actions (observer problem in epistemology). Find out info-computational mechanisms involved in DNA control of cellular processes.

*Answering questions for which natural computationalism is especially suitable framework,* such as: why is the genetic difference between humans and other animals smaller than we imagined before genome sequencing? How does the evident difference between humans and apes developed, given our social communication system as computational infrastructure that acts as a basis of human social intelligence established by natural computing? All of those questions can be framed in terms of information and natural computation and we are looking forward to see them addressed as the field of natural computing and study of computing nature develop.

From all above proposed research a richer notion of computation will emerge, which in its turn will help in the next step to better address natural phenomena as computations on informational structures. As Penrose in the Foreward to [15] states:

"(S)ome would prefer to define "computation" in terms of what a physical object can (in principle?) achieve (Deutsch, Teuscher, Bauer and Cooper). To me, however, this begs the question, and this same question certainly remains, whichever may be our preference concerning the use of the term "computation". If we prefer to use this "physical" definition, then all physical systems "compute" by definition, and in that case we would simply need a different word for the (original Church-Turing) mathematical concept of computation, so that the profound question raised, concerning the perhaps computable nature of the laws governing the operation of the universe can be studied, and indeed questioned."

It seems apt to conclude that nature indeed can be seen as a network of networks of computational processes and what we are trying is to compute the way nature does, learning its tricks of the trade. So the focus would not be *computability* but *computational modeling*. How good computational models of nature are we able to produce and what does it mean for a physical system to perform computation, computation being implementation of physical laws.

It is evident that natural computing/ computing nature presents a new natural philosophy of generality and scope that largely exceed natural philosophy of Newton's era, presented in his Philosophiae Naturalis Principia Mathematica. Natural computation brings us to the verge of a true paradigm shift in modeling, simulation and control of the physical world, and it remains to see how it will change our understanding of nature and especially living nature and humans, societies and ecologies.